# Coulomb and even-odd effects in cold and super-asymmetric fragmentation for thermal neutron induced fission of $^{235}$U

M. Montoya

Instituto Peruano de Energía Nuclear, Canadá 1470, San Borja, Lima, Peru
Universidad Nacional de Ingeniería, Av. Túpac Amaru 210, Rímac, Lima, Peru

**Abstract**

The Coulomb effects hypothesis is used to interpret even-odd effects of maximum total kinetic energy as a function of mass and charge of fragments from thermal neutron induced fission of $^{235}$U. Assuming spherical fragments at scission, the Coulomb interaction energy between fragments ($C_{\text{sph}}$) is higher than the $Q$-value, the available energy. Therefore at scission the fragments must be deformed, so that the Coulomb interaction energy does not exceed the $Q$-value. The fact that the even-odd effects in the maximum total kinetic energy as a function of the charge and mass, respectively, are lower than the even-odd effects of $Q$ is consistent with the assumption that odd mass fragments are softer than the even-even fragments.

Even-odd effects of charge distribution in super asymmetric fragmentation also are interpreted with the Coulomb effect hypothesis. Because the difference between $C_{\text{sph}}$ and $Q$ increases with asymmetry, fragmentations require higher total deformation energy to occur. Higher deformation energy of the fragments implies lower free energy to break pairs of nucleons. This explains why in the asymmetric fragmentation region, the even-odd effects of the distribution of proton number and neutron number increases with asymmetry. Based on a similar reasoning, a prediction of a relatively high even-odd effect in symmetric fragmentations is proposed.

## Introduction

The even-odd effects in the distribution of kinetic energy, charge and mass are among the properties that have generated controversy in the study of fragments from thermal neutron induced fission of actinides. To describe these effects is useful to recall some definitions related to them. Let be a fissile nucleus with charge $Z_f$ and mass $A_f$ that splits in a light fragment with $Z_L$ protons, $N_L$ neutrons (number of nucleons $A_L = Z_L + N_L$) and a heavy fragment with $Z_H$ protons, $N_H$ neutrons (number of nucleons $A_H = Z_H + N_H$). These numbers obey the following relations:

$$Z_f = Z_L + Z_H$$

and

$$A_f = A_L + A_H.$$

Based on these relations, to identify the two complementary fragments from a fission event it is enough to know the charge ($Z$) and the proton number ($N$) or the mass number ($A$) of the light fragment.

After scission, the light and heavy fragments acquire kinetic energies $K_L$, $K_H$ and excitation energies $X_L$, $X_H$, respectively.

Thus, the total kinetic energy ($K$) and the total excitation energy ($X$) are

$$K = K_\text{L} + K_\text{H}$$

and

$$X = X_\text{L} + X_\text{H},$$

respectively. These quantities are limited by the energy balance equation:

$$Q = K + X,$$

where $Q$ is the available energy of the reaction.

At the scission point, the available energy is spent into deformation energy ($D$), Coulomb interaction energy ($C$) and free energy ($F$):

$$Q = C + D + F.$$

The free energy is partitioned into intrinsic energy ($X^*$) and total pre-scission energy of fragments ($K_s$):

$$F = X^* + K_s.$$

The preference for even proton numbers in the fission fragments has been well established [1], leading to the definition of the even-odd effect of charge distribution ($\delta Z$):

$$\delta Z = \frac{Y_e^Z - Y_o^Z}{Y_e^Z + Y_o^Z}$$

where $Y_e^Z$ y $Y_o^Z$ are the yield of fragments with even and odd proton numbers, respectively. Similarly are defined the even-odd effect in the distribution neutron number ($\delta N$) and nucleon numbers ($\delta A$), respectively.

For a given fragmentation corresponding to proton and mass numbers $Z$ and $A$, respectively, the maximum total kinetic energy ($K_\text{max}$) is reached by a configuration that at scission acquires a maximum Coulomb interaction energy ($C_\text{max}$) and a minimum total deformation energy ($D_\text{min}$), limited by the equation

$$Q = C_\text{max} + D_\text{min}.$$

Because Coulomb repulsion between fragments, Coulomb interaction energy becomes total kinetic energy, so that:

$$K_\text{max} = C_\text{max} = Q - D_\text{min}.$$

Let be $A$ an odd nucleon number of the light fragment, the local even-odd effect in the maximum $Q$-value ($Q_\text{max}^A$) as a function of mass, is defined as

$$\delta_A Q_\text{max} = \frac{Q_\text{max}^{A-1} + Q_\text{max}^{A+1}}{2} - Q_\text{max}^A.$$

In average $\delta_A Q_\text{max}$ is positive. Similarly there are local even-odd effects in $Q_\text{max}$ as a function of proton number ($\delta_Z Q_\text{max}$) and neutron number ($\delta_N Q_\text{max}$), respectively.

Because the even-odd effect of charge and mass distribution, respectively, increases with the fragment kinetic energy [1], a positive local even-odd effect in the maximum total kinetic energy as a function of mass,

$$\delta_A K_\text{max} = \frac{K_\text{max}^{A-1} + K_\text{max}^{A+1}}{2} - K_\text{max}^A,$$

is expected. Similarly is expected positive values of $\delta_Z K_\text{max}$ and $\delta_N K_\text{max}$, which correspond to even-odd effects in the maximum total kinetic energy as a function of proton and neutron numbers, respectively.

Surprisingly, when C. Signarbieux et al. found the evidence of the existence of cold fission, in which the excitation energy is not enough for the fragments to emit neutrons, they do not find a significant even-odd effect in the distribution of the nucleon numbers ($\delta A \cong 0$). This set a controversy in those authors that, based on the even-odd effects in proton and neutron number distribution, respectively, supported the hypothesis that the fission process is superfluid [2]. However, in 1981, M. Montoya deduced [3, 4] that

$$\delta A = \delta Z + \delta N - 1,$$

which was confirmed by H. Nifenecker [5]. That relation shows that there is no contradiction between a null even-odd effect in $A$ distribution and no-null even-odd effects in $Z$ and $N$ distributions, respectively.

In 1991, based on data communicated by C. Signarbieux et al., F. Gönnenwein and B. Börsig show that the minimum excitation energy,

$$X_{\min} = Q_{K_{\max}} - K_{\max},$$

where ($Q_{K_{\max}}$) is the $Q$-value corresponding to the charge that maximizes the total kinetic energy, is lower for the odd than for the even proton numbers [6]. This result encourages research on even-odd effects, which leads to review the existent results and interpretations about even-odd effects in the distribution of mass, charge and maximum total kinetic energy of fragments.

**Even-odd effects in the maximum total kinetic energy**

Taking into account the $3 \times 10^6$ events from thermal neutron induced fission of $^{235}$U, obtained in 1981 by Signarbieux et al. [2], in 1984 M. Montoya presents the curve of threshold values for the 10 events with the highest total kinetic energy values [7]. These threshold values are assumed to correspond to the maximum total kinetic energy for each isobaric fragmentation. See Fig. 1. If one draws an straight line segments between the points corresponding to even nucleon numbers ($A_L$), the points corresponding to odd nucleon numbers are generally below that line, which is an indication that the average of $\delta_A K_{\max}$ is positive. Actually, the average value of $\delta_A K_{\max}$ is 0.5 MeV.

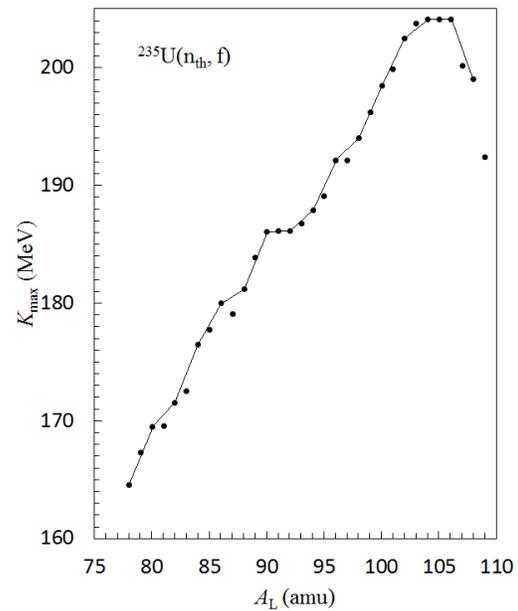

Fig. 1: Thermal neutron induced fission of $^{235}$U. Threshold values for which there are only 10 events with higher values of total kinetic energy of fragments. The total number of events is $3 \times 10^6$. Data taken from Ref. [7]. The average of even-odd effect on $K_{max}$ as a function of the fragments mass is 0.5 MeV.

In Fig. 2 one presents the maximum total kinetic energy as a function of charge and mass of fragments, obtained in 1986 by G. Simon et al. [8]. The cases where the charge that maximizes the total kinetic energy is odd are indicated. The average even-odd effect in the maximum total kinetic energy as a function of mass ($\delta_A K_{max}$) is 1 MeV. If we take groups of 3 masses corresponding to three different neighboring charges, the average of $\delta_Z K_{max}$ is = 0.66 MeV. Note that in these results for the masses 85 and 86, the charge that maximizes the total kinetic energy in both cases is 35.

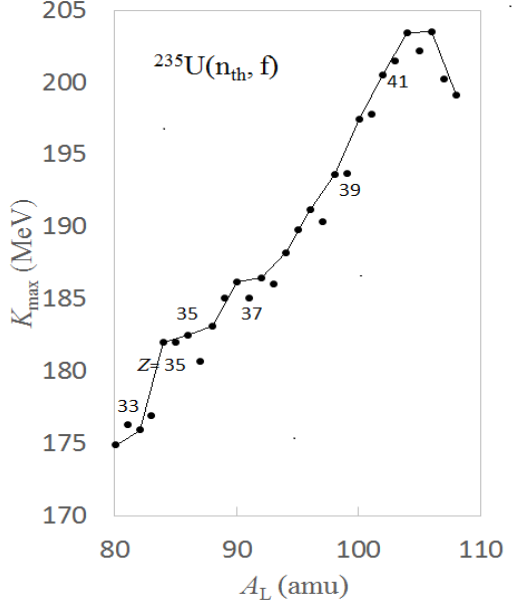

Fig. 2: Thermal neutron induced fission of $^{235}$U. Curve of the maximum total kinetic energy ($K_{max}$) as a function of the light fragment mass number is presented. Taken from Ref. [8]. The charges that maximize $K_{max}$ for each mass fragmentation are indicated.

In Fig. 3 the curve associated to the maximum total kinetic energy as a function of fragment mass communicated by C. Signarbieux is presented by F. Gönnenwein and B. Börsig [6]. In Fig. 4 one can see the charges maximizing the total kinetic energy and the charges that maximize the available energy, respectively. The $Q$-values are calculated using the mass table from Ref. [9]. The average of even-odd effect in the maximum total kinetic energy as a function of the mass ($\delta_A K_{max}$) is 0.5 MeV. If one takes 3 neighboring masses, corresponding to three different charges, in average it results $\delta_Z K_{max} = 0.8$ MeV. Note that the average of odd-even effect on the maximum available energy as a function of mass is 1.6 MeV.

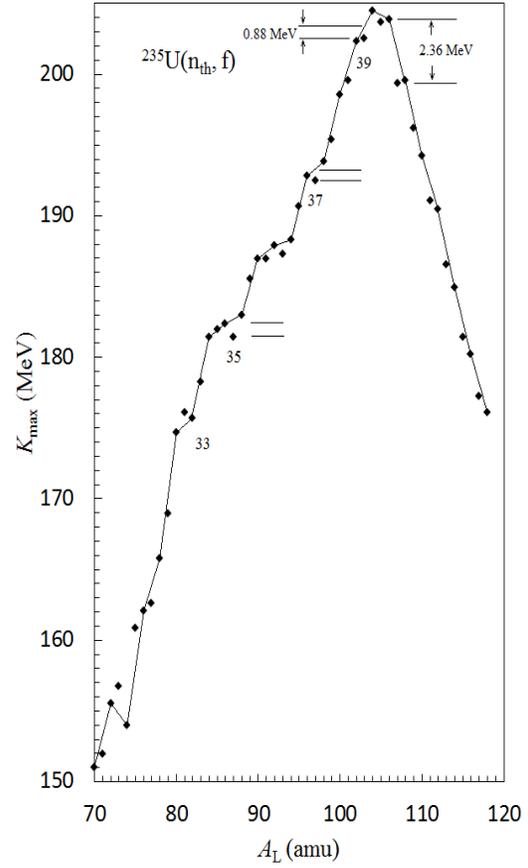

Fig. 3: Thermal neutron induced fission of $^{235}$U. Curve of the maximum total kinetic energy ($K_{max}$) as a function of the light fragment mass number is presented. Taken from Ref. [5]. The charges that maximize $K_{max}$ for each mass fragmentation are indicated.

To interpret this result one must remember that $Q_{K_{max}} = C_{max} + D_{min}$, regarding which it follows that

$$\delta_A C_{max} = \delta_A K_{max} = \delta_A Q_{K_{max}} - \delta_A D_{min},$$

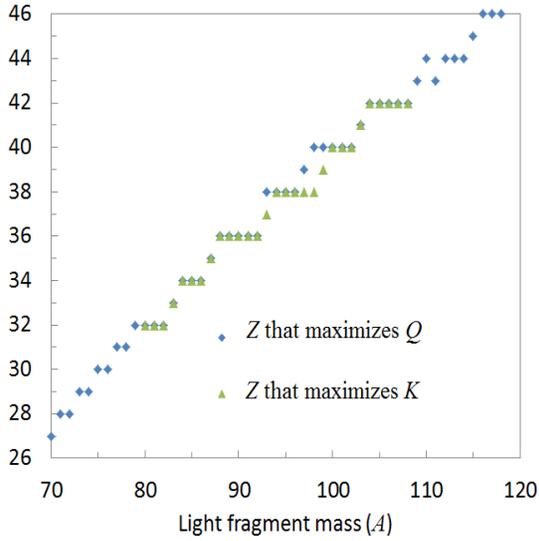

*Fig. 4: Thermal neutron induced fission of $^{235}U$. For each light fragment mass the charge that maximizes the Q-values (diamonds) and the charge maximizing the $K_{max}$ values (triangles) are presented. The Q-values are calculated using the mass table from Ref. [9]*

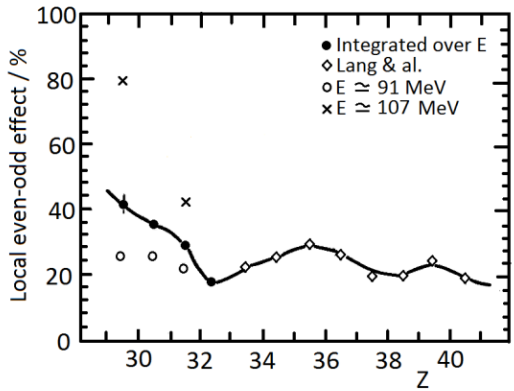

*Fig. 5: Thermal neutron induced fission of $^{235}U$. In the region of super-asymmetrical fission, the even-odd local effect on yields increases with the asymmetry of charge fragmentation. Taken from Ref. [13].*

The experimental result,

$$\delta_A K_{max} < \delta_A Q_{K_{max}},$$

implies that

$$\delta_A D_{min} > 0,$$

which suggests that the even-even fragments are harder than odd $A$ fragments, they need higher deformation energy to get the most compact configuration that obeys the relation $C_{max} = Q_{K_{max}} - D_{min}$.

In 2013, F. Gönnenwein and B. Börsig show that, for mass fragmentations 104/132, the kinetic energy associated to the charge fragmentation 41/51 reach the Q-value of the reaction, while the corresponding to the fragmentation 42/50 reaches a total maximum kinetic energy below 3 MeV the corresponding Q-value [10]. However, we should note that charge fragmentation 41/51 is more asymmetric than the 42/50 fragmentation. Therefore that result is also consistent with the Coulomb effect: for neighboring masses with similar values of energy available, the more asymmetric fragmentation reaches the higher values of total kinetic energy [11, 12].

In general, the results presented by F. Gönnenwein and B. Börsig show that the charges that maximize the total kinetic energy are the same as the charges that maximize the available energy. Exceptions occur for the masses 93, 96, 97 and 98, whose charges that maximize the available energy are 38, 39, 40 and 40, whereas the corresponding charges that maximize the total kinetic energy are 36 (<38), 38 (<39), 38 and 39 (<40), respectively. These results are also consistent with the hypothesis of the Coulomb effect [11, 12].

**Coulomb and even-odd effects in super-asymmetric fragmentations**

In 1989, J. L. Sida *et al.* note that in the region of super-asymmetric charge fragmentations, even-odd effect in charge and neutron number yields increase with charge asymmetry [13]. See Fig. 5. This result is consistent with the hypothesis of Coulomb effect. Indeed, if one assumes a scission configuration with spherical fragments whose surfaces, the Coulomb interaction energy is higher than the corresponding $Q$-value. Therefore at scission point the fragments must be deformed so that the energy of Coulomb interaction is lower than the available energy [11, 12]. For masses lower than 104 the Coulomb interaction energy curve separates from the value $Q$ to the extent that the fragmentation is asymmetric. This implies that the higher the asymmetry, the higher must be the deformation energy of the fission fragments will be to make the fission possible. Therefore, the process has lower free energy,

$$F = Q - C - D.$$

A lower free energy implies a lower probability to break pairs of nucleons, therefore a higher even-odd effect in yields of charge. This is precisely what experimentally is observed.

**Coulomb and even-odd effects in symmetric fragmentations**

Based on the Coulomb effect hypothesis, a prediction may be proposed. For symmetrical fragmentation, if one assumes scission configurations with spherical fragments, a higher separation of $C$ from $Q$ also occurs. Therefore, similarly to what happens with the asymmetric region, a greater even-odd effect may be expected. Indeed, there is a notorious even-odd effect in the curve of maximum kinetic energy for fragments in the mass region 104-108. See Fig. 3.

**Conclusion**

In cold fragmentation, the available energy is spent into Coulomb interaction energy and deformation energy, respectively. The fact that the even-odd effect in the maximum kinetic energy as a function of the mass or charge is lower than the even-odd effects in the available energy value suggests that, for the same deformation, even-even fragments need more energy than odd mass fragments.

Experimental results about cold fission suggest that, at scission, the total deformation energy of fragments competes with total intrinsic excitation energy: fragmentations which need high total deformation energy to occur will have low total intrinsic energy.

For the super-asymmetrical fragmentation the Coulomb interaction energy is much higher than the available energy. Those super-asymmetrical charge fragmentations need a super deformation to fulfill the energy balance condition; then a lower intrinsic excitation energy, which implies a lower probability for breaking of pairs of nucleons and, consequently, a higher even-odd effect of charge distribution is expected.

In the symmetric fission, the Coulomb interaction energy between supposedly spherical fragments is much higher than the $Q$-value. Then, similarly to the case of super-symmetric fragmentations, in symmetric region fragments should be quite deformed and, consequently, relatively high values of even-odd effects may be expected.